\documentclass[
aps,prd,
12pt,%10pt
nopreprintnumbers,
showpacs,
eqsecnum,
nofootinbib
]{revtex4-1}

\usepackage{graphicx}
\usepackage{amssymb}

\begin{document}

\title{Analytical Study of Charged Boson Stars with Large
Scalar Self-couplings
%\\ {\small ---A Poor Person's Approach to a Critical Phenomenon---}
}
\author{Nahomi Kan}\email[]{kan@gifu-nct.ac.jp}
\affiliation{National Institute of Technology, Gifu College,
Motosu-shi, Gifu 501-0495, Japan
}
\author{Kiyoshi Shiraishi}\email[]{shiraish@yamaguchi-u.ac.jp}
\affiliation{
Graduate School of Sciences and Technology for Innovation, Yamaguchi
University, Yamaguchi-shi, Yamaguchi 753--8512, Japan}
\date{\today}
%\date{}

\begin{abstract}
We give good approximate analytic solutions for
spherical charged boson stars in the large scalar-self-coupling limit in general
relativity. We show that if the charge $e$ and mass $m$ of the scalar field nearly
satisfy the critical relation $e^2\approx Gm^2$ (where $G$ is the Newton constant),
our analytic expressions for stable solutions
%qualitatively 
agree well with the numerical solutions. 
\end{abstract}

%\preprint{}


\maketitle

%%%%%%%%%%%%%%%%%%%%%%%%%%%%%%%%%%%%%%%%%%%%%%%%%%%%%%%%%%%%%%%%%%%%%%%%%%%
%Introduction
%%%%%%%%%%%%%%%%%%%%%%%%%%%%%%%%%%%%%%%%%%%%%%%%%%%%%%%%%%%%%%%%%%%%%%%%%%%
%%%%%%%%%%%%%%%%%%%%%%%%%%%%%%%%%%%%%%%%%%%%%%%%%%%%%%%%%%%%%%%%%%%%%%%%%%%
\section{Introduction}
\label{sec1}
%%%%%%%%%%%%%%%%%%%%%%%%%%%%%%%%%%%%%%%%%%%%%%%%%%%%%%%%%%%%%%%%%%%%%%%%%%%
%%%%%%%%%%%%%%%%%%%%%%%%%%%%%%%%%%%%%%%%%%%%%%%%%%%%%%%%%%%%%%%%%%%%%%%%%%%

One of the great problems in astroparticle physics
is the dark matter problem \cite{BHS,Profume}.
Many non-baryonic dark matter candidates have been supposed in the last several
decades; for instance, weakly interacting massive particles (WIMPs) have been
studied along with development of phenomenological supersymmetric particle theory.

There is a novel idea that condensation of unknown scalar bosons as a compact
object may play a role in dark matter.
Such a gravitating configuration is called a boson star \cite{Jetzer,LM,SM,LP} and
serves as a simple model to solve some problems arising in astrophysics, such as
galactic dynamics and stellar structure, avoiding restrictions on WIMPs and other
models.

Many authors have studied various models for boson stars so far,
and the studies on boson stars can yield
not only a clue to astrophysical problems but also
new insights into compact configurations in general relativity and in modified
gravity theories on theoretical grounds.

As a specific example, stable boson stars in scalar theory with a large
quartic self-coupling, first studied by Colpi et al.~\cite{colpi}, typically have a
large length scale, and thus the idea of boson stars with a galactic size naturally
arises as an explanation of the flat rotation curves of galaxies
\cite{LK,TCL,ST,ABBR,MA,BBAP,KS}.%
\footnote{The large values of couplings among unknown
fields may be compatible to the recently proposed scheme of strongly interacting
massive particles (SIMPs)
\cite{SIMP1,SIMP2}, though we need a hierarchical mass spectrum.}
 
Another interesting object is a charged boson star \cite{Jetzer,JB,JLS,DD,PQRR}.
The typical size of charged boson stars is larger than that of neutral boson
stars, because of partial subtraction of the magnitude of the attractive force
by the ``electric'' force.

Consider a system of particles with mass $m$ and charge $e$.
In the limit of $e^2\rightarrow Gm^2$, where $G$ is the Newton constant,
the long-range forces are mutually canceled.
%the Jeans length becomes infinity (at any temperature).
This fact raises a question: near the critical point $e^2=Gm^2$, can one find some
simple (or peculiar) behavior in the system?

A few decades ago, Jetzer {et al.} found the critical behavior in the mass of a
time-independent, spherical charged boson star \cite{Jetzer,JB}. Pugliese {et al.}
recently investigated such behavior for gravitating charged scalar theory
without scalar self-interactions
\cite{PQRR}. Both analyses relied on numerical methods. In the present paper, we
will study critical behaviors in stationary spherical charged boson stars
with a large scalar self-coupling, using analytical approximations.

If the charge is near critical, i.e., $e^2\approx Gm^2$, the equilibrium density
distribution is expected to be dilute as well as large scale.
The value of the central density goes to zero as the total mass becomes
infinite. Because the pressure in the center of the almost critical charged boson
star is small compared to the energy density, the configuration approaches a
Newtonian boson star in the critical limit.
Therefore, we first arrive at the idea of obtaining solutions for boson stars with
a small
$\epsilon$, which represents the (appropriately normalized) central value of a
square of the scalar field.

It should be noted that it is necessary to find solutions of the next order in
$\epsilon$, because the post-Newtonian effect determines the stability
of a definite mass and radius.
The boundary between a stable and an unstable star is given by the maximum mass.
It is reported \cite{Jetzer,JB} that the maximum mass increases with increasing
gauge coupling constant. An important aim of the present paper is to reproduce this
behavior semi-quantitatively in our approximation.

The present paper is organized as follows.
In Sec.~\ref{sec2}, we give the field equations for a boson star in general
relativity and their large coupling limit.
In Sec.~\ref{sec4}, in order to generate simple approximate solutions,
we construct an approximate differential equation for the square of the scalar
field. Linearizing the equation, we obtain analytical approximate solutions
expressed by trigonometric functions. The critical behavior in the mass of the
boson star is qualitatively confirmed by using the approximate solutions.
In Sec.~\ref{sec5}, we reconsider the energy density of the electric field, which
is ignored in Sec.~\ref{sec4}. After including the electromagnetic contribution to
the total mass, we again compare our approximate analysis with numerical results.
Finally, we summarize and discuss our results in
Sec.~\ref{so}.

In Appendix \ref{sec3}, a naive perturbative treatment of the equation as a power
expansion of $\epsilon$ is given. In Appendix \ref{ApB}, we present the
method which relies on the Taylor expansion in radius coordinates and estimates
the mass of the stable boson star in the critical limit.

%%%%%%%%%%%%%%%%%%%%%%%%%%%%%%%%%%%%%%%%%%%%%%%%
%%%%%%%%%%%%%%%%%%%%%%%%%%%%%%%%%%%%%%%%%%%%%%%%
\section{Charged boson stars in large coupling limit}
\label{sec2}
%%%%%%%%%%%%%%%%%%%%%%%%%%%%%%%%%%%%%%%%%%%%%%%%
%%%%%%%%%%%%%%%%%%%%%%%%%%%%%%%%%%%%%%%%%%%%%%%%
We consider an Einstein--Maxwell system with a self-interacting complex
scalar field
$\phi$ 
of  mass $m$ and charge $e$,
governed by the following action (where
$\hbar=c=1$):
\begin{eqnarray}
S&=&\int
d^4x\frac{\sqrt{-g}}{16\pi}\left[\frac{1}{G}R-F^2-
|D_\mu\phi|^2-m^2|\phi|^2-
\frac{{\lambda}}{2}|\phi|^4\right]\,,
\end{eqnarray}
where $d^4x=dt\,d^3\mbox{\boldmath $r$}$ , $G$ is the Newton constant,
$R$ is the scalar curvature, 
%$g$ is the determinant of the metric $g_{\mu\nu}$
%$(\mu,\nu=0,1,2,3)$.
and $F^2=g^{\mu\rho}g^{\nu\sigma}F_{\mu\nu}F_{\rho\sigma}$.
The field strength is defined as $F_{\mu\nu}=\partial_\mu A_\nu-\partial_\nu
A_\mu$, where $A_\mu$ is a $U(1)$ gauge field. The gauge field also appears in
$|D_\mu\phi|^2\equiv
g^{\mu\nu}(D_\mu\phi_i)^*(D_\nu\phi_i)$, where
the covariant derivative is $D_{\mu}=\partial_\mu+ie A_\mu$.
The scalar self-coupling constant ${\lambda}$ is assumed to be positive.

Varying the action with respect to the metric, we obtain the Einstein equation 
\begin{equation}
R^\mu_\nu-\frac{1}{2}\delta^\mu_\nu R=8\pi G T^\mu_\nu\,,
\label{EE}
\end{equation}
where the energy-momentum tensor $T_{\mu\nu}$ in the system is given by
\begin{eqnarray}
{16\pi}T_{\mu\nu}&=&(D_\mu\phi)^*(D_\nu\phi)+
(D_\nu\phi)^*(D_\mu\phi)-
g_{\mu\nu}|D_\rho\phi_i|^2-g_{\mu\nu}\Big(
m^2|\phi|^2
+\frac{{\lambda}}{2}|\phi|^4\Big)\nonumber \\
& &+4g^{\rho\sigma}F_{\rho\mu}F_{\sigma\nu}
-g_{\mu\nu}F^2
\,.
\end{eqnarray}
On the other hand, the equation of motion for the scalar field $\phi$ is given
by
\begin{equation}
D^\mu D_\mu \phi-m^2\phi-{\lambda}|\phi|^2\phi=0\,,
\label{KE}
\end{equation}
and the Maxwell equation is given by
\begin{equation}
D_\mu F^{\mu\nu}+i\frac{e}{4}[\phi^*D^\nu\phi-\phi(D^\nu\phi)^*]=0\,.
\label{ME}
\end{equation}

In the present paper, we consider stationary spherical boson stars. Thus, we assume
the metric with spherical symmetry
\begin{equation}
ds^2=-\left(1-\frac{2GM(r)}{r}\right)e^{-2\delta(r)}dt^2+
\left(1-\frac{2GM(r)}{r}\right)^{-1}dr^2+r^2(d\theta^2+\sin^2\theta d\varphi^2)\,.
\end{equation}
The ansatze for the scalar field and the gauge field are given by
\begin{equation}
\phi=\frac{1}{\sqrt{G}}\phi(r) e^{-i\omega t}\,,\quad
eA_\mu=m(A(r)+\omega)\delta_{\mu 0}\,,
\end{equation}
where $\omega$ is a constant.

Furthermore, we adopt the following new definitions of couplings:
\begin{equation}
q^2\equiv\frac{e^2}{Gm^2}\,,\quad
\Lambda\equiv\frac{\lambda}{Gm^2}\,.
\end{equation}

Substituting the foregoing ansatze and definitions in field equations (\ref{EE}),
(\ref{KE}), and (\ref{ME}) with the following replacement to
dimensionless variables
\begin{equation}
mr\rightarrow r\,,\quad GmM(r)\rightarrow M(r)\,,
\end{equation}
we obtain the simultaneous differential equations
\begin{eqnarray}
& &\phi''+\left(\frac{2}{r}-\delta'+\frac{2}{r}\frac{M-rM'}{r-2M}\right)\phi'+
\left[\frac{e^{2\delta}}{1-\frac{2M}{r}}A^2-1-\Lambda\phi^2
\right]\frac{1}{1-\frac{2M}{r}}\phi=0\,,\\
& &A''+\left(\frac{2}{r}+\delta'\right)A'-\frac{q^2}{2}
\frac{1}{1-\frac{2M}{r}}\phi^2A=0\,,\\
& &\frac{2}{r^2}M'=e^{2\delta}\frac{{A'}^2}{q^2}+\frac{1}{2}
\left[\frac{e^{2\delta}}{1-\frac{2M}{r}}A^2\phi^2+
\left(1-\frac{2M}{r}\right){\phi'}^2\right]
+\frac{1}{2}\phi^2+\frac{\Lambda}{4}\phi^4\,,\\
& &-\frac{1}{r}\delta'=\frac{1}{2}
\left[\frac{e^{2\delta}}{\left(1-\frac{2M}{r}\right)^2}A^2\phi^2+
\phi'^2\right]\,,
\end{eqnarray}
where the prime (${}'$) indicates the derivative with respect to $r$.

To realize the configuration of a spherical boson star, we impose the
following boundary conditions:
\begin{equation}
\phi'(0)=0\,, \quad A'(0)=0\,,\quad
M(0)=0\,,\quad\delta(0)=0\,,\quad \phi(\infty)=0\,,\quad A(\infty)=const.
\end{equation}
Note that an arbitrary value for $\delta(0)$ is allowed because it can be
absorbed by the redefinition of the time coordinate $t$.

Here, we consider the large coupling limit \cite{Jetzer,colpi}.
%It is incidentally known that the large coupling leads 
%to a large-scale boson star. 
To take the limit, we introduce the following quantities:
\begin{equation}
x\equiv\frac{g}{\sqrt{\Lambda}}r\,,\quad
\Phi(x)\equiv\sqrt{\Lambda}\phi(r)\,,\quad
\mu(x)\equiv\frac{gM(r)}{\sqrt{\Lambda}}\,,
\label{scale}
\end{equation}
where
\begin{equation}
g=\sqrt{1-q^2}\,.
\end{equation}
Owing to new variables, we can take the limit of $\Lambda\rightarrow\infty$;
the field equations then become \cite{Jetzer}
\begin{eqnarray}
& &\sigma\equiv\Phi^2=\frac{e^{2\delta}}{1-\frac{2\mu}{x}}A^2-1\,,
\label{eqeqs}\\
& &A''+\left(\frac{2}{x}+\delta'\right)A'-\frac{q^2}{2g^2}
\frac{1}{1-\frac{2\mu}{x}}\sigma A=0\,,
\label{eqeqa}\\
& &\frac{2}{x^2}\mu'=e^{2\delta}\frac{{A'}^2}{q^2}+\frac{1}{2}
\frac{e^{2\delta}}{1-\frac{2\mu}{x}}A^2\frac{\sigma}{g^2}
+\frac{1}{2}\frac{\sigma}{g^2}+\frac{1}{4}\frac{\sigma^2}{g^2}\,,
\label{eqeqm}\\
& &-\frac{1}{x}\delta'=\frac{1}{2}
\frac{e^{2\delta}}{\left(1-\frac{2\mu}{x}\right)^2}A^2\frac{\sigma}{g^2}\,,
\label{eqeqd}
\end{eqnarray}
where, and hereafter, the prime (${}'$) stands for $\frac{d}{dx}$.
Note that the first equation shows an algebraic relation valid for
$\sigma=\Phi^2>0$. Therefore, the field equations are now reduced to three
differential equations on $A(x)$, $\mu(x)$, and $\delta(x)$.
The surface of the spherical boson star is defined by the radius $x=x_*$ at which
$\sigma(x_*)=\Phi^2(x_*)=0$. Outside the boson star, it is thought that $\Phi(x)$
vanishes for $x>x_*$.

The numerical solutions for the system in the large coupling limit have been
investigated, for example, in Refs.~\cite{Jetzer,JB}.
We will consider an approximation that leads to analytic solutions for the system
in the next section.

%%%%%%%%%%%%%%%%%%%%%%%%%%%%%%%%%%%%%%%%%%%%%%%%%%%%%%%%%%%%%%%%%%%%%%%%%%%
%%%%%%%%%%%%%%%%%%%%%%%%%%%%%%%%%%%%%%%%%%%%%%%%%%%%%%%%%%%%%%%%%%%%%%%%%%%
\section{Approximate equation for  square of the scalar field}
\label{sec4}
%%%%%%%%%%%%%%%%%%%%%%%%%%%%%%%%%%%%%%%%%%%%%%%%%%%%%%%%%%%%%%%%%%%%%%%%%%%
%%%%%%%%%%%%%%%%%%%%%%%%%%%%%%%%%%%%%%%%%%%%%%%%%%%%%%%%%%%%%%%%%%%%%%%%%%%
When solving the field equations mathematically, we initially regard the
region of definition for $A(x)$, $\mu(x)$, and $\delta(x)$ as $[0,\infty)$,
though physical meanings of the solutions hold only in the region of positive
$\sigma(x)=\Phi^2(x)$; i.e., $[0, x_*]$.

Here, we again give the field equations for $\alpha(x)\equiv
A(x)/A(0)\equiv A(x)/A_0$,
$\mu(x)$, and
$\delta(x)$ ((\ref{eqeqa}), (\ref{eqeqm}), and (\ref{eqeqd})):
\begin{eqnarray}
& &\alpha''(x)+\left(\frac{2}{x}+\delta'(x)\right)\alpha'(x)-\frac{q^2}{2g^2}
\frac{1}{1-\frac{2\mu}{x}}\sigma(x)\alpha(x)=0\,,
\label{ae}\\
& &\frac{2}{x^2}\mu'(x)=e^{2\delta(x)}A_0^2\frac{{\alpha'(x)}^2}{q^2}
+\frac{1}{g^2}\sigma(x)\left(1+\frac{3}{4}\sigma(x)\right)\,,
\label{me}\\
& &-\frac{1}{x}\delta'(x)=\frac{1}{2g^2}
\frac{1}{1-\frac{2\mu}{x}}{\sigma(x)(1+\sigma(x))}
\label{de}\,,
\end{eqnarray}
where
\begin{equation}
\sigma(x)=\frac{e^{2\delta(x)}}{1-\frac{2\mu(x)}{x}}A_0^2
\alpha^2(x)-1\,.
\label{ss}
\end{equation}

%%%
If $q^2\approx 1$, an attractive force (gravity) and
a repulsive force (Coulomb repulsion) almost cancel each
other out. Under the restriction that the particle number is constant, the density
becomes low (because of the repulsive force which originates from the
self-interaction of scalars\footnote{Even if the self-interaction is absent, the
density is expected to become low due to the uncertainty principle (``quantum
force'').}) for a stable boson star.
%%%

Therefore, we can examine the expansion in terms of a ``small'' parameter
$\epsilon$ defined by $\epsilon\equiv{A_0^2-1}=\sigma(0)$ for solving the
differential equations. Although the equations at the lowest order of $\epsilon$
become very simple, those at the next order are very complicated to analyze.
This approach is described in Appendix \ref{sec3}.
Thus, in the present section, we consider the other approach.

Now, we investigate the relationship in terms of derivatives of $\sigma(x)$ by
using the following approximation. 
Because we wish to consider a stable dilute boson star, we first assume 
$\frac{2\mu(x)}{x}\ll 1$ and
$\delta(x)\ll 1$, which are near-vacuum values of
the variables. We will, however, take care of their derivative, which is
expressed by
$\sigma(x)$ and its derivative.

We next assume 
$(\alpha'(x))^2\ll\sigma(x)/g^2$. This assumption implies that the energy density
of the electric field is negligible compared with the energy density of the scalar
field and can therefore be omitted, because the right-hand side of Eq.~(\ref{me})
is proportional to the total energy density. For finite values of $\sigma(x)$ and
small $g^2$, this approximation is reasonable. Because we are now going to study
the behavior of
$\sigma(x)$ in the nearly flat background, we take only the first order in the
electric field in the present
approximation.  It is noteworthy that we do not assume $\sigma(x)\ll 1$ at the
first time, whose value should be small for a stable dilute boson star in the
critical limit $g\approx 0$. 

Then, by using Eq.~(\ref{ss}), we can approximate the first derivative of
$\ln(1+\sigma(x))$ as
\begin{equation}
\frac{\sigma'(x)}{1+\sigma(x)}\approx
2\frac{\alpha'(x)}{\alpha(x)}+\frac{2\mu'(x)}{x}-\frac{2\mu(x)}{x^2}+2\delta'(x)\,.
\label{s1}
\end{equation}
Under the same assumptions, the field equations (\ref{me}) and (\ref{de})
can be interpreted as
\begin{eqnarray}
& &\frac{2}{x}\mu'(x)\approx
\frac{x}{g^2}\sigma(x)\left(1+\frac{3}{4}\sigma(x)\right)\,,
\label{masseq}\\ &
&2\delta'(x)\approx-\frac{x}{g^2}
{\sigma(x)(1+\sigma(x))}\,,
\label{deltaeq}
\end{eqnarray}
where we find that $\mu'(x)$ and $\delta'(x)$ are supposed to be determined by
$\sigma(x)$. Eq.~(\ref{s1}) then becomes
\begin{equation}
\frac{\sigma'(x)}{1+\sigma(x)}\approx
2\frac{\alpha'(x)}{\alpha(x)}-\frac{x}{4g^2}\sigma^2(x)-\frac{2\mu(x)}{x^2}\,.
\label{s11}
\end{equation}
One more differentiation of the above equation yields
\begin{eqnarray}
\frac{\sigma''(x)}{1+\sigma(x)}&\approx&
2\frac{\alpha''(x)}{\alpha(x)}-2\left(\frac{\alpha'(x)}{\alpha(x)}\right)^2
-\frac{1}{4g^2}\sigma^2(x)-\frac{x}{2g^2}\sigma(x)\sigma'(x)
-\frac{2\mu'(x)}{x^2}+\frac{4\mu(x)}{x^3}\nonumber \\
& &+
\left(\frac{\sigma'(x)}{1+\sigma(x)}\right)^2\nonumber \\
&\approx&
2\frac{\alpha''(x)}{\alpha(x)}-\frac{1}{g^2}\sigma(x)(1+\sigma(x))-\frac{x}{2g^2}\sigma(x)\sigma'(x)
+\frac{4\mu(x)}{x^3}\nonumber \\
& &+
\left(\frac{\sigma'(x)}{1+\sigma(x)}\right)^2
-2\left(\frac{\alpha'(x)}{\alpha(x)}\right)^2\,,
\label{s2}
\end{eqnarray}
where we used Eq.~(\ref{masseq}).
Combining Eqs.~(\ref{s11}) and (\ref{s2}), we obtain 
\begin{eqnarray}
& &\frac{1}{1+\sigma(x)}\left(\sigma''(x)+\frac{2}{x}\sigma'(x)\right)\nonumber \\
& &\approx
2\left(\frac{\alpha''(x)}{\alpha(x)}+\frac{2}{x}\frac{\alpha'(x)}{\alpha(x)}\right)
-\frac{1}{g^2}\sigma(x)\left(1+\frac{3}{2}\sigma(x)\right)-
\frac{x}{2g^2}\sigma(x)\sigma'(x)\nonumber \\
& &\quad+
\left(\frac{\sigma'(x)}{1+\sigma(x)}\right)^2
-2\left(\frac{\alpha'(x)}{\alpha(x)}\right)^2\nonumber
\\ & &\approx
-2\delta'(x)\frac{\alpha'(x)}{\alpha(x)}+\frac{q^2}{g^2}\sigma(x)-
\frac{1}{g^2}\sigma(x)\left(1+\frac{3}{2}\sigma(x)\right)-
\frac{x}{2g^2}\sigma(x)\sigma'(x)\nonumber \\
& &\quad+
\left(\frac{\sigma'(x)}{1+\sigma(x)}\right)^2
-2\left(\frac{\alpha'(x)}{\alpha(x)}\right)^2\nonumber
\\ & &\approx
-\sigma(x)\left(1+\frac{3}{2g^2}\sigma(x)\right)
-2\delta'(x)\frac{\alpha'(x)}{\alpha(x)}-
\frac{x}{2g^2}\sigma(x)\sigma'(x)\nonumber \\
& &\quad+
\left(\frac{\sigma'(x)}{1+\sigma(x)}\right)^2
-2\left(\frac{\alpha'(x)}{\alpha(x)}\right)^2\,,
\label{eqeq}
\end{eqnarray}
where we used an approximate equation that comes from Eq.~(\ref{ae}),
\begin{equation}
\alpha''(x)+\left(\frac{2}{x}+\delta'(x)\right)\alpha'(x)-\frac{q^2}{2g^2}
\sigma(x)\alpha(x)=0\,.
\label{eqe}
\end{equation}
Note also that $\frac{q^2-1}{g^2}=-1$.

The second and third terms in Eq.~(\ref{eqeq}) are reduced, if we can
further approximate
$\frac{\sigma'(x)}{1+\sigma(x)}$ by $2\frac{\alpha'(x)}{\alpha(x)}$ and use
Eq.~(\ref{deltaeq}), to
\begin{eqnarray}
& &-2\delta'(x)\frac{\alpha'(x)}{\alpha(x)}-
\frac{x}{2g^2}\sigma(x)\sigma'(x)\nonumber \\
& &\approx
-2\delta'(x)\frac{\alpha'(x)}{\alpha(x)}-
\frac{x}{2g^2}\sigma(x)(1+\sigma(x))\frac{\sigma'(x)}{1+\sigma(x)}\nonumber \\
& &\approx
-2\delta'(x)\frac{\alpha'(x)}{\alpha(x)}
+2\delta'(x)\frac{\alpha'(x)}{\alpha(x)}\nonumber
\\ & &\approx 
0\,.
\label{eqf}
\end{eqnarray}

Now, if we assume $(\alpha'(x))^2\ll\sigma(x)/g^2$,
$\frac{2\mu(x)}{x}\ll 1$, 
$\delta(x)\ll 1$, and adopt the
additional assumption $|\frac{x}{4g^2}\sigma^2(x)+\frac{2\mu(x)}{x^2}|\ll
2\frac{\alpha'(x)}{\alpha(x)}$, the following differential equation on
$\sigma(x)$ holds:
\begin{equation}
\sigma''(x)+\frac{2}{x}\sigma'(x)+(1+\sigma(x))
\left[\sigma(x)+\frac{3}{2g^2}\sigma^2(x)-\frac{1}{2}
\left(
\frac{\sigma'(x)}{1+\sigma(x)}\right)^2\right]=0\,.
\end{equation}
Unfortunately, exact solutions for this nonlinear equation are not known.
Although it is interesting to solve this nonlinear equation, 
we here consider a further approximation to solve the equation analytically.%
\footnote{A comment is given in Sec.~\ref{so}.}

In order to solve the equation approximately, we first set
$\sigma(x)\equiv\epsilon\tilde{\sigma}(0)$. We then find
\begin{equation}
\tilde{\sigma}''(x)+\frac{2}{x}\tilde{\sigma}'(x)+(1+\epsilon\tilde{\sigma}(x))
\left[\tilde{\sigma}(x)+\frac{3\epsilon}{2g^2}\tilde{\sigma}^2(x)-
\frac{\epsilon^2(\tilde{\sigma}'(x))^2}{2\left(1+\epsilon\tilde{\sigma}(x)\right)^2}
\right]=0\,.
\label{req1}
\end{equation}
The boundary condition at $x=0$ then should be
$\tilde{\sigma}(0)=1$ and $\tilde{\sigma}'(0)=0$.

As an approximation scheme, we further assume that the value of $\epsilon$, which
indicates the central value of the scalar density, is small for a stable dilute
boson star. We then find the following approximate linear equation:
\begin{equation}
\tilde{\sigma}''(x)+\frac{2}{x}\tilde{\sigma}'(x)+(1+\epsilon)
\left(1+\frac{3\epsilon}{2g^2}\right)\tilde{\sigma}(x)=0\,.
\label{req2}
\end{equation}
%%%
This approximation is considered to be good, especially for $x\approx 0$, where
$\tilde{\sigma}\approx 1$ and $\tilde{\sigma}'\approx 0$. 
For a stable dilute boson star, it is expected that the behavior of the solution
near the origin determines its overall shape and size.
Furthermore, we have left the ``next-leading'' terms in $\epsilon$ because
the limit of $\epsilon\rightarrow 0$ in Eq.~(\ref{req1}) or Eq.~(\ref{req2})
yields the result corresponding to the Newtonian limit and,
as previously stated, we wish to study the relativistic mass of the
critically charged boson star. 
%%%

The solution for the above linearized equation satisfying $\tilde{\sigma}(0)=1$ is
\begin{equation}
\tilde{\sigma}(x)=\frac{\sin kx}{kx}\,,
\label{sinek}
\end{equation}
where
\begin{equation}
k=k(\epsilon)\equiv\sqrt{(1+\epsilon)
\left(1+\frac{3\epsilon}{2g^2}\right)}\,.
\end{equation}

The approximate equation for $\mu(x)$ (\ref{masseq}) then yields
\begin{equation}
\mu(x)\approx\frac{\epsilon}{2g^2k^3}\left[\sin kx- kx\cos
kx+\frac{3\epsilon}{8}\left(kx-\frac{\sin 2kx}{2}\right)\right]\,.
\end{equation}

\begin{figure}[ht]
\centering
\includegraphics[width=7cm]{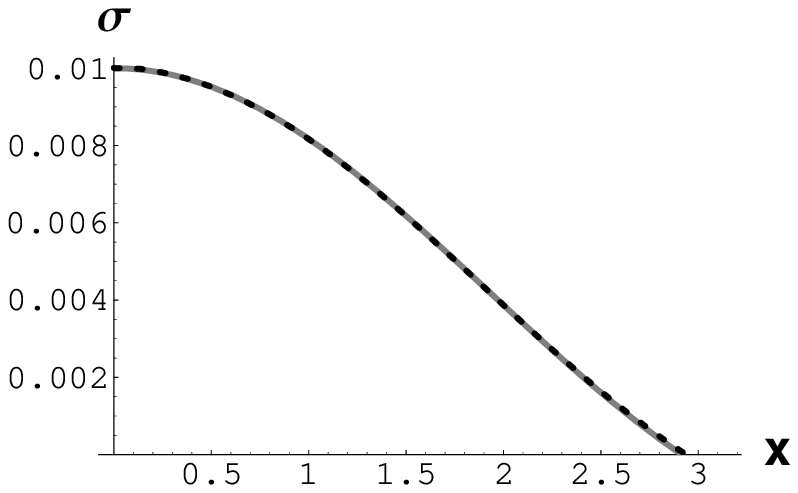}
\includegraphics[width=7cm]{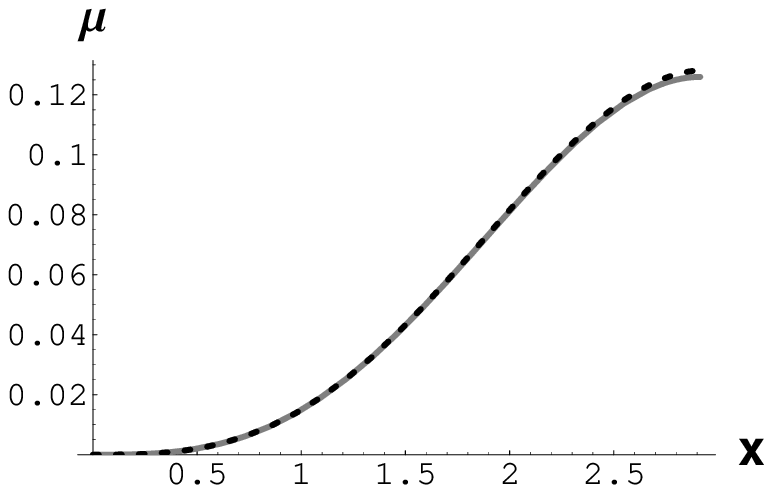}
\caption{
Numerical solutions in the large coupling limit (dotted lines) and the
present approximation (light solid lines) for $\sigma(x)$ and $\mu(x)$ on the
interval $[0,
\pi/k]$, where
$g^2=0.1$ and
$\epsilon=0.01$.}
\label{figs21}
\end{figure}

\begin{figure}[ht]
\centering
\includegraphics[width=7cm]{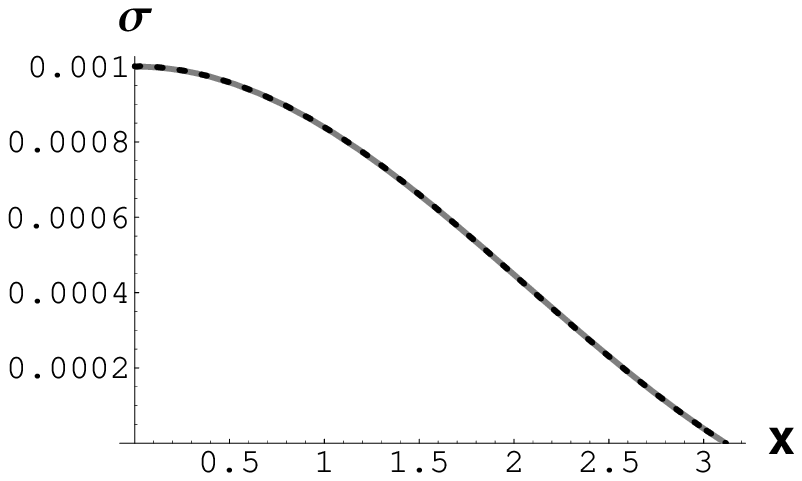}
\includegraphics[width=7cm]{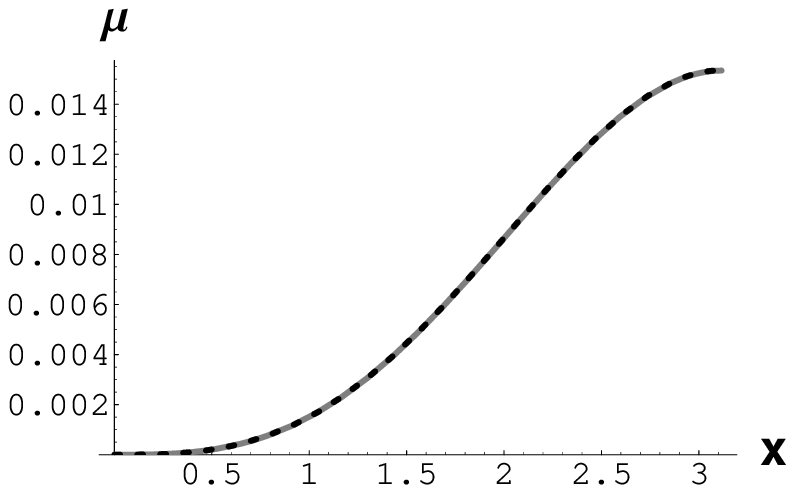}
\caption{
Numerical solutions in the large coupling limit (dotted lines) and the present
approximation (light solid lines) for $\sigma(x)$ and $\mu(x)$ on the
interval $[0,
\pi/k]$, where $g^2=0.1$
and
 $\epsilon=0.001$.}
\label{figs31}
\end{figure}

In Fig.~\ref{figs21} (for $g^2=0.1$ and
$\epsilon=0.01$) and Fig.~\ref{figs31} (for $g^2=0.1$ and
$\epsilon=0.001$), dotted lines are the numerical results in the large coupling
limit, and light solid lines are our approximations.
One can see that they almost coincide.

%%%%%%%%%%%%%%%%%%%%%%%%%%%
% 21
%%%%%%%%%%%%%%%%%%%%%%%%%%%
%\begin{wrapfigure}{r}{5cm}
%\end{wrapfigure}
%%%%%%%%%%%%%%%%%%%%%%%%%%%

%%%%%%%%%%%%%%%%%%%%%%%%%%%
% 31
%%%%%%%%%%%%%%%%%%%%%%%%%%%
%\begin{wrapfigure}{r}{5cm}
%\end{wrapfigure}
%%%%%%%%%%%%%%%%%%%%%%%%%%%

The surface of the boson star is located at $x=x_*$ where
$\sigma(x_*)=0$. By using our approximate solution (\ref{sinek}),
we easily find that  $x_*(\epsilon)=\pi/k$ and
\begin{equation}
\mu(x_*(\epsilon))\approx\frac{\epsilon\pi}{2g^2k^3(\epsilon)}\left(1+\frac{3\epsilon}{8}
\right)\,.
\end{equation}

\begin{figure}[ht]
\centering
\includegraphics[width=7cm]{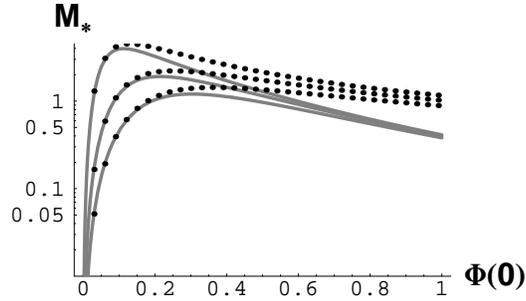}
\caption{
Approximate charged boson star mass $M_*$ in units of $\sqrt{\Lambda}/(Gm)$ as
a function of
$\Phi(0)=\sqrt{\Lambda}\phi(0)$ for $g=0.1, 0.2, \mbox{~and~} 0.3$ (from the upper
line to the lower line) for the case $\Lambda\rightarrow\infty$. The dots indicate
the numerical results in the large coupling limit.}
\label{figms}
\end{figure}

Our definition of mass in the present section is the integration of the energy
density inside a boson star, where $\sigma(x)>0$.%
\footnote{The discussion on the definition of mass corresponding to that in
Ref.~\cite{Jetzer,JB} is given in Sec.~\ref{sec5}.}
That is
\begin{equation}
M_*\equiv\frac{\mu(x_*(\epsilon))}{g}=\frac{GmM(x_*(\epsilon))}{\sqrt{\Lambda}}\,,
\end{equation}
and we show $M_*$ as a function of $\Phi(0)=\sqrt{\epsilon}$ in Fig.~\ref{figms}.
In this figure, dots indicate the numerical dependence of the mass with respect to
$\Phi(0)$ for $g=0.1, 0.2, \mbox{~and~} 0.3$, whereas light solid curves represent
our approximation for
$g=0.1, 0.2, \mbox{~and~} 0.3$. 

$M_*$ reaches its maximum value as $\Phi(0)$ increases.
The behaviors of the approximate values look alike as in figure (Fig.~9) in
Ref.~\cite{Jetzer} for small
$\Phi(0)$, whereas, unfortunately, they look different  for large $\Phi(0)$ (as
expected, because the present approximation relies only on small
$\Phi(0)=\sqrt{\epsilon}$).

%%%%%%%%%%%%%%%%%%%%%%%%%%%
% ms
%%%%%%%%%%%%%%%%%%%%%%%%%%%
%\begin{wrapfigure}{r}{5cm}
%\end{wrapfigure}
%%%%%%%%%%%%%%%%%%%%%%%%%%%

In the general relativistic system, it is known that the mass of the star
increases monotonically up to a maximum as the central density increases.
The maximum mass defines the border between the stable and unstable configurations.

To obtain solutions for the maximum boson star mass, the value of
$\epsilon=\sigma(0)=\Phi^2(0)$ is the root of the equation
\begin{equation}
\frac{\partial\mu(x_*(\epsilon))}{\partial\epsilon}=0\,.
\end{equation}
Unfortunately, this equation reduces to a third-order equation in $\epsilon$.
We do not have to solve the equation so precisely  beyond the present
approximation scheme. Thus, dropping the third-order term in the equation for
sufficiently small
$\epsilon$, we obtain the approximate solution
\begin{equation}
\epsilon=\epsilon_m\approx \frac{4}{3}g^2\frac{6}{3-g^2+\sqrt{9+168g^2-11g^4}}\,.
\end{equation}
Further, if we consider the limit of
$g\rightarrow 0$, we find $\epsilon_m\rightarrow\frac{4}{3}g^2$,
$k(\epsilon_m)\rightarrow \sqrt{3}$, and
\begin{equation}
\lim_{g\rightarrow 0}\mu(x_*(\epsilon_m))\approx\frac{2\sqrt{3}\pi}{27}=0.403\,.
\end{equation}
Thus, the mass of the stable charged boson star is found to be
\begin{equation}
M_{*max}\approx 0.403\frac{1}{\sqrt{1-q^2}}\,,
\end{equation}
for a small $g=\sqrt{1-q^2}$.
For a small $g$,
\begin{equation}
\Phi(0)_{*max}\approx\sqrt{\frac{4}{3}}g=1.15{\sqrt{1-q^2}}\,,
\end{equation}
and
\begin{equation}
r_{*max}\equiv
\frac{x_*(\epsilon_m)}{g}\approx\frac{\pi}{g
k(\epsilon_m)}\approx\frac{\pi}{g\sqrt{3}}=1.81\frac{1}{\sqrt{1-q^2}}\,.
\end{equation}

Jetzer and Bij \cite{Jetzer,JB} gave (in our notation) $\Phi(0)_{*max}\approx
(2.43/ 2^{3/4})\sqrt{1-q^2}=1.44\sqrt{1-q^2}$,
$M_{*max}\approx 0.226\times
2^{3/4}\times\sqrt{2}\frac{1}{\sqrt{1-q^2}}=0.537\frac{1}{\sqrt{1-q^2}}$.%
\footnote{The factor $2^{3/4}$ comes from the different definition of critical
charge and its power, and the factor $\sqrt{2}$ comes from the different
definition of $\lambda/G$.} Therefore, the deviation from the precise value is
$\sim\! 33\%$ for
$M_{*max}$ and
$\sim\! 20\%$ for $\Phi(0)_{*max}$.
The definition of the radius of the boson star in Ref.~\cite{Jetzer} is the average
of
$r_*$  over the particle density. In the present approximation, the function
$\sigma(x)$ represents both the particle and the energy density. Thus, their
definition of the radius should be recognized as
$R_*\approx 0.344~r_{*max}$, because the solution for $\sigma(x)$ is proportional
to
$\sin kx/kx$, and
$\{\int_0^\pi x(\sin x/x) dx\}/\{\pi\int_0^\pi (\sin x/x) dx\}\approx
0.344$. Our approximate value is
$R_*=0.415\times
2^{3/4}\times\sqrt{2}\frac{1}{\sqrt{1-q^2}}=0.985\frac{1}{\sqrt{1-q^2}}
=0.344\times 2.86\frac{1}{\sqrt{1-q^2}}$. The deviation of $r_{*max}$ is
considered to be $\sim\! 58\%$.

Finally, in this section, we mention that it is also possible to show the
qualitative
$M_*$-$\sqrt{\epsilon}$ relation
in the other approximation. 
The approximation, which utilizes the approximate functions of the order $O(x^2)$,
is shown in Appendix \ref{ApB}.

In the next section, we reconsider the definition of mass and inclusion of the
energy density of the electric field.

%%%%%%%%%%%%%%%%%%%%%%%%%%%%%%%%%%%%%%%%%%%%%%%%%%%%%%%%%%%%%%%%%%%%%%%%%%%
%%%%%%%%%%%%%%%%%%%%%%%%%%%%%%%%%%%%%%%%%%%%%%%%%%%%%%%%%%%%%%%%%%%%%%%%%%%
\section{The energy of the electric field}
\label{sec5}
%%%%%%%%%%%%%%%%%%%%%%%%%%%%%%%%%%%%%%%%%%%%%%%%%%%%%%%%%%%%%%%%%%%%%%%%%%%
%%%%%%%%%%%%%%%%%%%%%%%%%%%%%%%%%%%%%%%%%%%%%%%%%%%%%%%%%%%%%%%%%%%%%%%%%%%
In the approximation scheme in the last section,
we assumed $(\alpha')^2\ll\sigma/g^2$.
This corresponds to the omission of the energy density of the electric field
as it is negligible compared with the energy density of the scalar field.

In the present section, we estimate the contribution of the electric energy
density to the boson star mass.
Although accounting for the electric energy in addition to the scalar energy
sounds inconsistent judging from the ansatz,
it can be considered that the configuration is the main source field of
all the fields
because the approximation for the scalar field
configuration
$\sigma(x)$ fits numerical computations very well.
Thus, we insist that the addition of the electric energy density has a physical
meaning.

The definition of mass in Refs.~\cite{Jetzer,JB} includes the contribution of
the electric field.
It is pointed out \cite{PQRR} that $M_*$ (in our notation) differs from
the ``actual'' mass (which is proportional to the coefficient of the inverse of
the distance from the origin in the asymptotic region). The difference is due to
the electric contribution and has been ignored in the approximation scheme in
Sec.~\ref{sec4}.

We adopt Eq.~(\ref{eqe}), but we omit the term $\alpha'\delta'$ in the
equation, where the term is smaller than the source term $\propto \sigma$. We
then get
\begin{equation}
\alpha(x)\approx 1+\frac{q^2\epsilon}{2g^2k^2}\left(1-\frac{\sin kx}{kx}\right)
\qquad\mbox{for } x<x_*=\pi/k\,.
\label{51}
\end{equation}
Note that in this section we consider the solutions for $\alpha(x)$ inside the
boson star $(x<x_*)$ and outside the boson star $(x>x_*)$ separately. Therefore,
the electric contribution to the mass inside the boson star can be estimated as
\begin{equation}
\Delta
M_*(\epsilon)=\frac{1+\epsilon}{2g}\int_0^{x_*}\frac{(\alpha'(x))^2}{q^2}dx=\frac{\pi
q^2\epsilon^2(1+\epsilon)}{16g^5k^5}\,,
\end{equation}
where we approximated $\delta\sim 0$.
The ratio of the correction
\begin{equation}
\frac{\Delta
M_*(\epsilon)}{M_*(\epsilon)}=\frac{
q^2\epsilon(1+\epsilon)}{8g^2k^2\left(1+\frac{3\epsilon}{8}\right)}
\end{equation}
is at most $8\%$ in the parameter region of Fig.~\ref{figms}, and the
approximate value of $M_*+\Delta M_*$ cannot be larger than that of the numerical
result.

Next, we consider the contribution of the electric field outside the boson star.
At the surface of the boson star, $x=x_*=\pi/k$, and from Eq.~(\ref{51}), we have
\begin{equation}
\alpha(x_*)\approx\frac{q^2\epsilon}{2g^2k^2}\,,
\quad\alpha'(x_*)\approx\frac{
q^2\epsilon}{2g^2k\pi}\,.
\end{equation}

Outside the boson star, $\alpha$ behaves in
accordance with $\alpha(x)=C_0+\frac{C_1}{x}$, which is the solution of the field
equation $\alpha''(x)+\frac{2}{x}\alpha'(x)=0$. Thus, we find the following
solution for $\alpha(x)$ outside the boson star, which is smoothly connected to
the solution for $\alpha$ inside the boson star:
\begin{equation}
\alpha(x)\approx\frac{q^2\epsilon}{2g^2k^2}-\frac{\pi
q^2\epsilon}{2g^2k^3x}\qquad\mbox{for }x>x_*\,.
\end{equation}
Using this solution, we find the electric energy outside the charged boson star
as
\begin{equation}
\Delta M_{outside}=\frac{1+\epsilon}{2g}\int_{x_*}^\infty
e^{2\delta(x_*)}\frac{(\alpha'(x))^2}{q^2}\,.
\end{equation}
In the present approximation scheme,
$\delta(x_*)=-\frac{\epsilon}{g^2k^2}+O(\epsilon^2)$. Therefore, we again ignore
$\delta(x_m)$ and obtain
\begin{equation}
\Delta M_{outside}(\epsilon)=\frac{\pi q^2\epsilon^2
(1+\epsilon)}{8g^5k^5}\,,
\end{equation}
which is the same order as $\Delta M_*$.

\begin{figure}[ht]
\centering
\includegraphics[width=7cm]{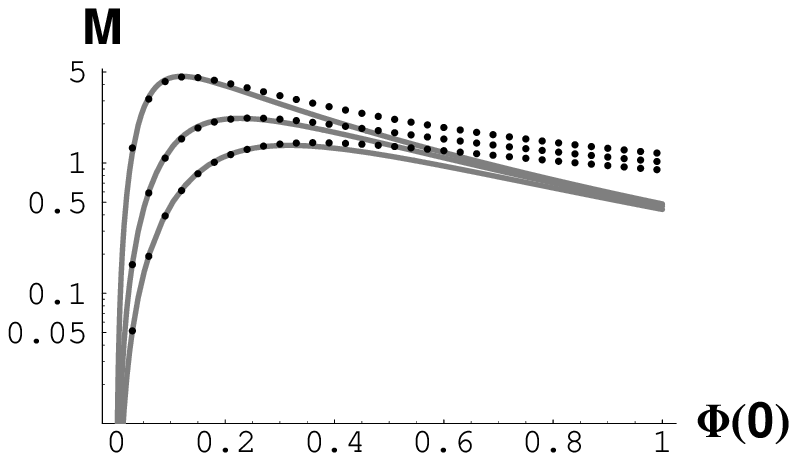}
\caption{
Approximate charged boson star mass $M$ (including the electric energy) in
units of
$\sqrt{\Lambda}/(Gm)$ as a function of
$\Phi(0)=\sqrt{\Lambda}\phi(0)$ for $g=0.1, 0.2,\mbox{~and~} 0.3$ (from the upper
line to the lower line) for the case $\Lambda\rightarrow\infty$. The dots indicate
the numerical results in the large coupling limit.}
\label{figmall}
\end{figure}

We then estimate the total mass as
\begin{equation}
M(\epsilon)\equiv M_*(\epsilon)+\Delta M_*(\epsilon)+\Delta
M_{outside}(\epsilon)=\frac{\epsilon\pi}{2g^3k^3}\left(1+\frac{3\epsilon}{8}
\right)+\frac{3\pi q^2\epsilon^2 (1+\epsilon)}{16g^5k^5}\,,
\end{equation}
which is illustrated in Fig.~\ref{figmall} for $g=0.1,0.2,\mbox{~and~} 0.3$.
For small values of $\Phi(0)=\sqrt{\epsilon}$, approximate values fit
the numerical results  better than the previous approximation.

%%%%%%%%%%%%%%%%%%%%%%%%%%%
% mall
%%%%%%%%%%%%%%%%%%%%%%%%%%%
%\begin{wrapfigure}{r}{5cm}
%\end{wrapfigure}
%%%%%%%%%%%%%%%%%%%%%%%%%%%

The maximum mass $M$ is attained if $\frac{\partial
M(\epsilon)}{\partial\epsilon}=0$, which reduces to
\begin{equation}
\frac{\epsilon}{g^2}\rightarrow\frac{4}{15}(2+\sqrt{14})=1.53
\qquad \mbox{for } \epsilon, g^2\rightarrow 0\,.
\end{equation}
The maximum mass is then
\begin{equation}
M_{max}\approx\frac{10\sqrt{5}(11+3\sqrt{14})\pi}{3(\sqrt{7}+\sqrt{2})^{5}g}
=0.472\frac{1}{\sqrt{1-q^2}}\,.
\end{equation}
The deviation of $M$ from the values in Refs.~\cite{Jetzer,JB} is now approximately
$12\%$, while $\Phi(0)_{max}\approx\sqrt{1.53}g=1.24\sqrt{1-q^2}$, and the
deviation is
$14\%$. We have now obtained good approximate values by including the electric
energy contribution.

%%%%%%%%%%%%%%%%%%%%%%%%%%%%%%%%%%%%%%%%%%%%%%%%%%%%%%%%%%%%%%%%%
%%%%%%%%%%%%%%%%%%%%%%%%%%%%%%%%%%%%%%%%%%%%%%%%%%%%%%%%%%%%%%%%%
\section{Summary and discussion}
\label{so}
%%%%%%%%%%%%%%%%%%%%%%%%%%%%%%%%%%%%%%%%%%%%%%%%%%%%%%%%%%%%%%%%%
%%%%%%%%%%%%%%%%%%%%%%%%%%%%%%%%%%%%%%%%%%%%%%%%%%%%%%%%%%%%%%%%%

In this paper, we presented  approximate solutions for
dilute charged boson stars with spherical symmetry in the large scalar
self-coupling limit. 
An approximation scheme is presented in Sec.~\ref{sec4}, where
we first consider the approximate differential equation for the square of the
scalar field $\sigma(x)$. In this approximation, we assumed that the contribution
of the energy density of the electric field is relatively small. A further
linearized approximation yields a fully analytic approximation for a charged
boson star. In Sec.~\ref{sec5}, we improved the approximation by reconsidering the
electric energy.
Because it has been recognized that solutions with an $\epsilon$ value that
is smaller than the maximum
$\epsilon_m$ value are stable and the others are unstable,
our approximation has a certain physical meaning for stable configurations of
charged boson stars.

We confirmed that the maximum mass of the boson star increases with the gauge
coupling constant as
$(\sqrt{1-q^2})^{-1}$ for a charge close to the critical charge $e^2\approx
Gm^2$ in our approximation, whose deviation from the numerical result is on
the order of a few ten percent.

It was pointed out that there is a localized configuration even if the
charge of the scalar field is larger than the critical coupling for
scalar theory without self-coupling \cite{PQRR}.
The analysis of the critical behavior of the maximum mass in the large
self-coupling limit under consideration is nevertheless valid, because the large
coupling limit does not yield higher node solutions \cite{Jetzer}, whereas only
solutions with nodes exist for over-critical cases as reported in Ref.~\cite{PQRR}
for scalar theory with no self-interaction.

Our analytically approximate solutions can be used to check the validity of
numerical solutions generally. Analytic solutions can also be used as a background
configuration in an investigation of the quantum vacuum around charged boson stars
\cite{JLS}, as well as the seeds of an exact solution (for instance, nonspherical)
in numerical computations.

We would like to improve the approximation for not so small $\epsilon$.
To this end, we have to try a basic approach such as the Pad\'e approximation.
In Sec.~\ref{sec4}, a nonlinear equation for $\sigma(x)$ has been derived.
We wish to use some type of renormalization group methods
\cite{CGO1,CGO2,BB1,BB2} to evaluate the solution, though it is difficult to
directly apply the known methods to the present form of the equation.

Finally, we should consider the analysis of charged boson stars in scalar theory
with an arbitrary self-coupling. We hope to return to these and other subjects in
future work.

%%%%%%%%%%%%%%%%%%%%%%%%%%%%%%%%%%%%%%%%%%%%%%%%%%%%%%%%%%%%%%%%%
%%%%%%%%%%%%%%%%%%%%%%%%%%%%%%%%%%%%%%%%%%%%%%%%%%%%%%%%%%%%%%%%%
\appendix
%%%%%%%%%%%%%%%%%%%%%%%%%%%%%%%%%%%%%%%%%%%%%%%%%%%%%%%%%%%%%%%%%
%%%%%%%%%%%%%%%%%%%%%%%%%%%%%%%%%%%%%%%%%%%%%%%%%%%%%%%%%%%%%%%%%
%%%%%%%%%%%%%%%%%%%%%%%%%%%%%%%%%%%%%%%%%%%%%%%%%%%%%%%%%%%%%%%%%%%%%%%%%%%
%%%%%%%%%%%%%%%%%%%%%%%%%%%%%%%%%%%%%%%%%%%%%%%%%%%%%%%%%%%%%%%%%%%%%%%%%%%
\section{Perturbative expansion in terms of $\epsilon$}
\label{sec3}
%%%%%%%%%%%%%%%%%%%%%%%%%%%%%%%%%%%%%%%%%%%%%%%%%%%%%%%%%%%%%%%%%%%%%%%%%%%
%%%%%%%%%%%%%%%%%%%%%%%%%%%%%%%%%%%%%%%%%%%%%%%%%%%%%%%%%%%%%%%%%%%%%%%%%%%
Here, we solve the field equations obtained in
Sec.~\ref{sec2} as a perturbative expansion in $\epsilon$.
First, we define 
\begin{eqnarray}
A(x)&=&
A_0\alpha(x)=A_0(1+\epsilon \alpha_1(x)+\epsilon^2
\alpha_2(x)+\cdots)
\,,\label{exp1}\\
\mu(x)&=&\epsilon \mu_1(x)+\epsilon^2 \mu_{2}(x)+\cdots\,,
\label{exp2}\\
\delta(x)&=&\epsilon \delta_{1}(x)+\epsilon^2 \delta_{2}(x)+\cdots\,,
\label{exp3}
\end{eqnarray}
where $\epsilon$ is a parameter defined as
\begin{equation}
\epsilon\equiv A_0^2-1\quad\mbox{or}\quad A_0\equiv\sqrt{1+\epsilon}\,,
\end{equation}
which means
\begin{equation}
\sigma(0)=\Phi^2(0)=\epsilon\,,
\end{equation}
if we identify boundary conditions $\alpha_i(0)=\mu_i(0)=\delta_i(0)=0$
$(i=1, 2, \dots)$.
The value of $\epsilon$ is expected to be small in the critical limit
$q^2\approx 1$ or equivalent to $g^2\approx 0$.

Substituting the above series expansions (\ref{exp1}), (\ref{exp2}), and
(\ref{exp3}) into the field equations (\ref{eqeqs}), (\ref{eqeqa}), (\ref{eqeqm}),
and (\ref{eqeqd}), we find the equations in the first order in powers of
$\epsilon$, 
\begin{eqnarray}
\alpha_1''+\frac{2}{x}\alpha_1'&=&\frac{q^2}{g^2}\left(\frac{1}{2}+\alpha_1+
\frac{\mu_{1}}{x}+\delta_1\right)
\,,\\
\frac{1}{x^2}\mu_{1}'&=&\frac{1}{g^2}
\left(\frac{1}{2}+\alpha_1+\frac{\mu_{1}}{x}+\delta_1\right)\,,\\
\frac{1}{x}\delta_1'&=&-\frac{1}{g^2}\left(\frac{1}{2}+\alpha_1+
\frac{\mu_{1}}{x}+\delta_1\right)\,,
\end{eqnarray}
which are just the linearized field equations.

In this order, 
using $\alpha_1''+\frac{2}{x}\alpha_1'=\frac{1}{x^2}(x^2\alpha_1')'
=\frac{1}{x}(x\alpha_1)''$ and paying attention to the similarity of the
right-hand sides of the equations,
we can easily obtain analytic
solutions under the boundary conditions at $x=0$ as
\begin{eqnarray}
\alpha_1(x)&=&\frac{q^2}{2g^2}\left(1-\frac{\sin x}{x}\right)
\,,\\
\mu_{1}(x)&=&\frac{1}{2g^2}
\left(\sin x-x\cos x\right)\,,\\
\delta_1(x)&=&-\frac{1}{2g^2}\left(1-\cos x\right)\,.
\end{eqnarray}

Up to this order, the profile of the scalar field is found to be
\begin{eqnarray}
\Phi^2(x)&\approx& (1+\epsilon)(1+2\epsilon\delta_1(x))
\left(1+\epsilon\frac{2\mu_{1}(x)}{x}\right)(1+\epsilon\alpha_1(x))^2-1\nonumber \\
&=&\epsilon\left(1+2\alpha_1(x)+\frac{2\mu_{1}(x)}{x}+2\delta_1(x)\right)
+O(\epsilon^2)\nonumber \\
&=&\epsilon\frac{\sin x}{x}+O(\epsilon^2)\,.
\end{eqnarray}

This profile has been obtained in the same system by the Newtonian approximation.
Because we now treat dilute boson stars, the result is just a verification of the
present lowest-order analysis.

The field equations in the second order of $\epsilon$ can be read as
\begin{eqnarray}
\alpha_2''+\frac{2}{x}\alpha_2'&=&\frac{q^2}{g^2}\left(%\frac{\rho}{2}+
\alpha_2+
\frac{\mu_{2}}{x}+\delta_2\right)+f_\alpha
\,,\\
\frac{1}{x^2}\mu_{2}'&=&\frac{1}{g^2}
\left(%\frac{\rho}{2}+
\alpha_2+\frac{\mu_{2}}{x}+\delta_2\right)+f_\mu\,,\\
\frac{1}{x}\delta_2'&=&-\frac{1}{g^2}\left(%\frac{\rho}{2}+
\alpha_2+
\frac{\mu_{2}}{x}+\delta_2\right)+f_\delta\,,
\end{eqnarray}
where
\begin{eqnarray}
f_\alpha&=&-\alpha_1'\delta_1'+\frac{q^2}{g^2}\left[
\left(\alpha_1+\frac{2\mu_{1}}{x}\right)
\left(\frac{1}{2}+\alpha_1+\frac{\mu_{1}}{x}+\delta_1\right)
+X\right]
\,,\\
f_\mu&=&\frac{1}{2q^2}\alpha_1'^2+\frac{1}{g^2}\left[
\frac{3}{2}
\left(\frac{1}{2}+\alpha_1+\frac{\mu_{1}}{x}+\delta_1\right)^2
+X\right]\,,\\
f_\delta&=&-\frac{1}{g^2}\left[
\left(1+2\alpha_1+\frac{4\mu_{1}}{x}+2\delta_1\right)
\left(\frac{1}{2}+\alpha_1+\frac{\mu_{1}}{x}+\delta_1\right)
+X\right]\,,\\
X&=&\alpha_1+\frac{\mu_{1}}{x}+\delta_1+
\frac{\alpha_1^2}{2}+2\frac{\mu_{1}^2}{x^2}+\delta_1^2+
\frac{2\mu_{1}}{x}\alpha_1+\frac{2\mu_{1}}{x}\delta_1+2\alpha_1\delta_1\,.
\end{eqnarray}

These inhomogeneous differential equations can be solved easily.
For this purpose, we have only to know that the inhomogeneous equation
\begin{equation}
u''(x)+u(x)=f(x)
\end{equation}
has a general solution (where $A$ and $B$ are integration constants)
\begin{equation}
u(x)=A\sin x+B\cos x+\sin x\int_0^xf(t)\cos t\, dt-\cos x\int_0^xf(t)\sin t\, dt\,.
\end{equation}
The solutions are given by
\begin{eqnarray}
\alpha_2&=&\frac{q^2}{8g^4}\Big\{-[\gamma_E+\ln 2x-\mbox{Ci}(2x)]
-(1-\cos x)\frac{\sin x}{x}%+(4\rho
-3%)
g^2\left(1-\frac{\sin x}{x}\right)\nonumber \\
& &+\frac{3}{4}(3+g^2)\left[(\ln
3+\mbox{Ci}(x)-\mbox{Ci}(3x))\frac{\sin
x}{x}-\frac{3\mbox{Si}(x)-\mbox{Si}(3x)}{x}\cos x\right]\Big\}\,,\\
\mu_{2}&=&-\frac{1}{8g^4}\Big\{x\sin^2x+\frac{1}{2}(3-g^2)(2x-\sin 2x)
+(1-g^2)\left(\frac{\sin^2 x}{x}-x\right)\nonumber \\ & &+(1+%(1-4\rho)
g^2)(\sin
x-x\cos x)\nonumber \\ & &+\frac{3}{4}(3+g^2)[(\ln
3+\mbox{Ci}(x)-\mbox{Ci}(3x))(\sin x-x\cos x)\nonumber \\
& &-
(3\mbox{Si}(x)-\mbox{Si}(3x))(\cos x+x\sin x)]\Big\}\,,\\
\delta_2&=&\frac{1}{8g^4}\Big\{3\sin^2 x+(1+%(1-4\rho)
g^2)(1-\cos
x)+(1-g^2)[\gamma_E+\ln 2x-\mbox{Ci}(2x)]\nonumber \\
& &-\frac{3}{4}(3+g^2)\left[(\ln
3+\mbox{Ci}(x)-\mbox{Ci}(3x))\cos x+(3\mbox{Si}(x)-\mbox{Si}(3x))\sin
x\right]\Big\}
\,,
\end{eqnarray}
where $\gamma_E$ is the Euler--Mascheroni constant,
the sine integral is defined as $\mbox{Si}(z)\equiv\int_0^z\frac{\sin t}{t}dt$,
and the cosine integral is defined as $\mbox{Ci}(z)\equiv-\int_z^\infty\frac{\cos
t}{t}dt$.
Note that the mathematical relations
$\gamma_E+\ln 2x-\mbox{Ci}(2x)=\int_0^{2x}\frac{1-\cos t}{t}dt$
and
$\ln 3+\mbox{Ci}(x)-\mbox{Ci}(3x)=\int_0^{3x}\frac{1-\cos t}{t}dt
-\int_0^{x}\frac{1-\cos t}{t}dt$ have been used.

%%%%%%%%%%%%%%%%%%%%%%%%%%%
% 21
%%%%%%%%%%%%%%%%%%%%%%%%%%%
%\begin{wrapfigure}{r}{5cm}
\begin{figure}[ht]
\centering
\includegraphics[width=5cm]{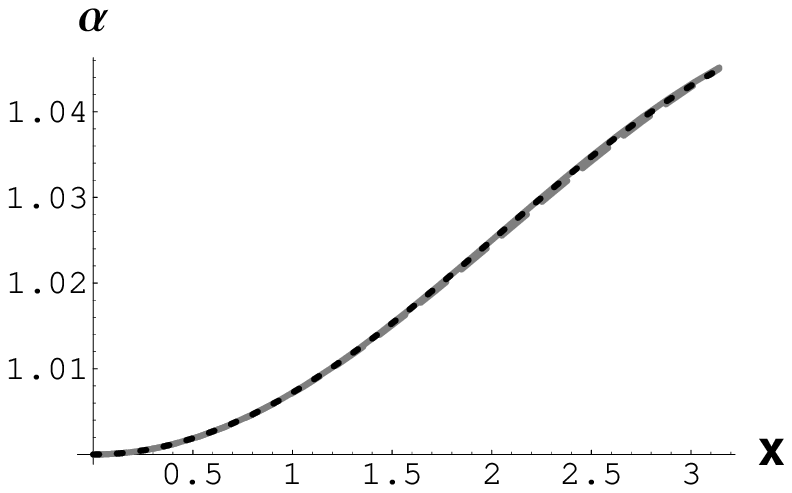}
\includegraphics[width=5cm]{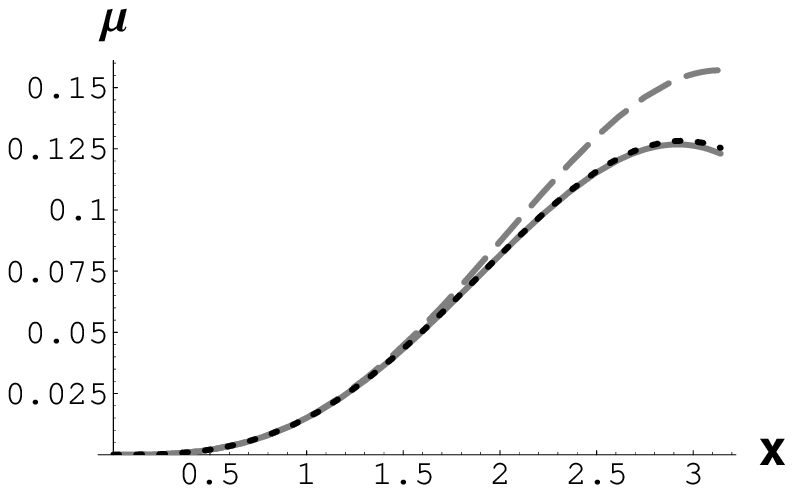}
\includegraphics[width=5cm]{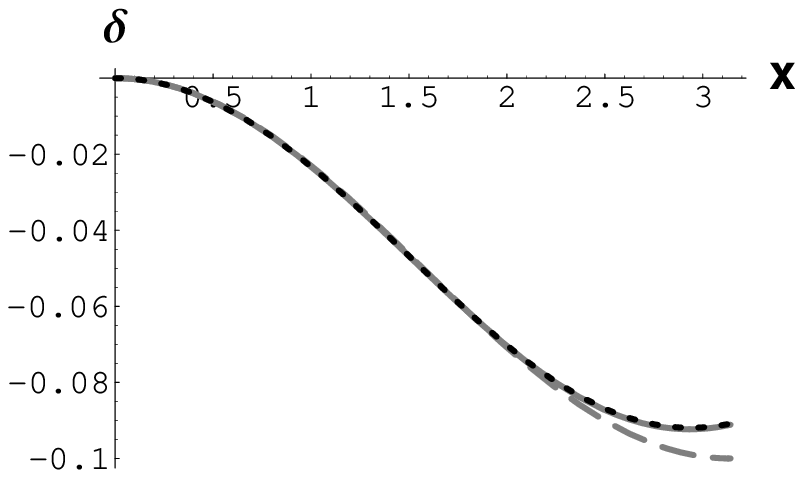}
\caption{
Numerical solutions in the large coupling limit (dotted lines) and perturbative
approximations (light dashed lines for the first-order approximations in $\epsilon$
and light solid lines for the second-order approximation in $\epsilon$) for
$\alpha(x)$,
$\mu(x)$, and
$\delta(x)$ on the
interval $[0,
\pi]$, where $g^2=0.1$ and $\epsilon=0.01$.}
\label{fig21}
\end{figure}
%\end{wrapfigure}
%%%%%%%%%%%%%%%%%%%%%%%%%%%

%%%%%%%%%%%%%%%%%%%%%%%%%%%
% 31
%%%%%%%%%%%%%%%%%%%%%%%%%%%
%\begin{wrapfigure}{r}{5cm}
\begin{figure}[ht]
\centering
\includegraphics[width=5cm]{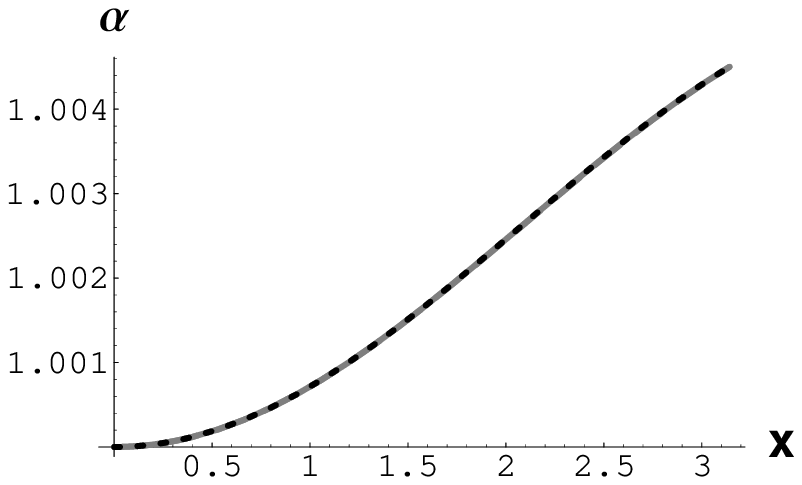}
\includegraphics[width=5cm]{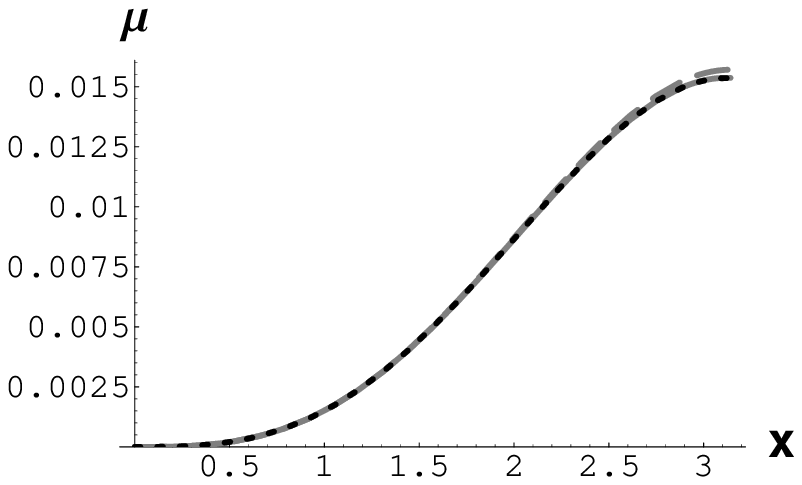}
\includegraphics[width=5cm]{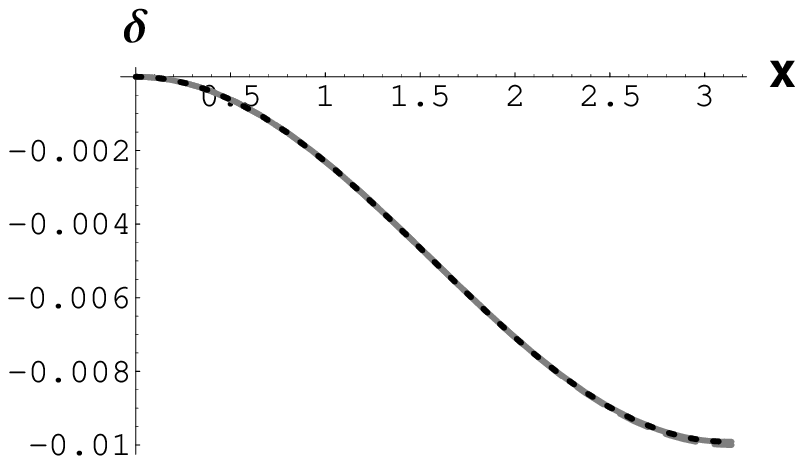}
\caption{
Numerical solutions in the large coupling limit (dotted lines) and perturbative
approximations (light dashed lines for the first-order approximations in $\epsilon$
and light solid lines for the second-order approximation in $\epsilon$) for
$\alpha(x)$,
$\mu(x)$, and
$\delta(x)$ on the
interval $[0,
\pi]$, where $g^2=0.1$ and $\epsilon=0.001$.}
\label{fig31}
\end{figure}
%\end{wrapfigure}
%%%%%%%%%%%%%%%%%%%%%%%%%%%

The perturbative solutions in comparison with numerical calculations in the
large coupling limit (Eqs.~(\ref{eqeqs}), (\ref{eqeqa}), (\ref{eqeqm}) and
(\ref{eqeqd})) are exhibited in Fig.~\ref{fig21} (for
$g^2=0.1$ and
$\epsilon=0.01$) and Fig.~\ref{fig31} (for $g^2=0.1$ and $\epsilon=0.001$).  
The approximation is good for small values of $\epsilon$, as expected.
We also find that near the critical charge $g\sim 0$ $(q\sim 1)$,
the $i$-th order functions seems $\sim\! O((\epsilon/g^2)^i)$.
Thus, if $\epsilon/g^2\ll1$, the perturbative approximation works well
and the solutions up to the second order agree with the numerical results. 

The solutions are parameterized by
 $\epsilon=\sigma(0)$.
The stability of the objects described by the solutions is discussed
by variations of this parameter.%
\footnote{In some cases, however, quasi-stable configurations with a very long
lifetime may be admitted as astrophysical objects.} 
It is
demonstrated that the configuration is stable if the boson star mass takes
the maximum value with respect to variations of this parameters.

To obtain the maximum mass of boson stars, we first evaluate the value of
$\mu(x)/g$ at the surface of the boson star $x=x_*$ and next consider the
variation with respect to the parameter $\epsilon$.
The perturbed solution we obtained is not suitable for such calculations in
the analytic method, because of the complexity seen in the second-order solutions.

%%%%%%%%%%%%%%%%%%%%%%%%%%%%%%%%%%%%%%%%%%%%%%%%%%%%%%%%%%%%%%%%%%%%%%%%%%%
%%%%%%%%%%%%%%%%%%%%%%%%%%%%%%%%%%%%%%%%%%%%%%%%%%%%%%%%%%%%%%%%%%%%%%%%%%%
\section{Approximation by the Taylor expansion in $x$}
\label{ApB}
%%%%%%%%%%%%%%%%%%%%%%%%%%%%%%%%%%%%%%%%%%%%%%%%%%%%%%%%%%%%%%%%%%%%%%%%%%%
%%%%%%%%%%%%%%%%%%%%%%%%%%%%%%%%%%%%%%%%%%%%%%%%%%%%%%%%%%%%%%%%%%%%%%%%%%%
We consider the solution  for the field
equations  (\ref{ae})--(\ref{de}) at the lowest order in the Taylor expansion in
$x$ around $x=0$ with the boundary condition $\sigma(0)=\epsilon$.
We then find
\begin{eqnarray}
\alpha(x)&\approx&1+\frac{q^2\epsilon}{12 g^2}x^2\,,\\
\frac{\mu(x)}{x}&\approx&\frac{\epsilon(3\epsilon+4)}{24 g^2}x^2\,,\\
\delta(x)&\approx&-\frac{\epsilon(1+\epsilon)}{4 g^2}x^2\,.
\end{eqnarray}
Accordingly, using Eq.~(\ref{ss}), we find $\sigma(x)$ at the lowest order as
\begin{eqnarray}
\sigma(x)&\approx& (1+\epsilon)(1+2\delta(x))
\left(1+\frac{2\mu(x)}{x}\right)(\alpha(x))^2-1\nonumber \\
&\approx&\epsilon\left(1-\frac{3\epsilon+2g^2}{12g^2}x^2\right)\,,
\end{eqnarray}
where we assume that $\epsilon$ is small, which is expected for a stable dilute
boson star. We obtain the radius of the boson star
from the present approximation as
\begin{equation}
x_*(\epsilon)\approx\sqrt{\frac{12g^2}{3\epsilon+2g^2}}\,,
\end{equation}
and
\begin{equation}
\mu(x_*(\epsilon))\approx\frac{g\epsilon(3\epsilon+4)}{24}
\left(\frac{12}{3\epsilon+2g^2}\right)^{3/2}\,.
\end{equation}

\begin{figure}[ht]
\centering
\includegraphics[width=7cm]{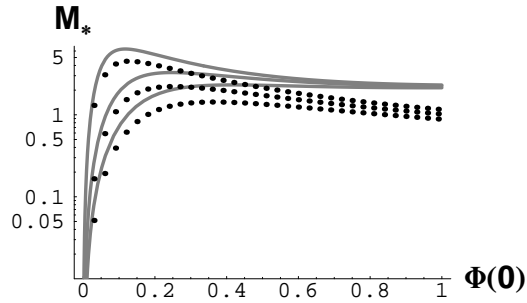}
\caption{
Charged boson star mass (approximated using the second-order functions) in units of
$\sqrt{\Lambda}/(Gm)$ as a function of
$\Phi(0)=\sqrt{\Lambda}\phi(0)$ for $g=0.1, 0.2,\mbox{~and~} 0.3$ (from the upper
line to the lower line) for the case $\Lambda\rightarrow\infty$. The dots indicate
the numerical results.}
\label{figmt}
\end{figure}

We show $M_*=\mu(x_*(\epsilon))/g$ as a function of $\sqrt{\epsilon}=\Phi(0)$
in Fig.~\ref{figmt}. The qualitative behavior of the graph is similar to that in
Refs.~\cite{Jetzer,JB} (see also the discussion in Sec.~\ref{sec4}).

%%%%%%%%%%%%%%%%%%%%%%%%%%%
% mt
%%%%%%%%%%%%%%%%%%%%%%%%%%%
%\begin{wrapfigure}{r}{5cm}
%\end{wrapfigure}
%%%%%%%%%%%%%%%%%%%%%%%%%%%

The maximum mass for a fixed $g$ is given with $\epsilon$ that satisfies the
following equation:
\begin{equation}
\frac{\partial \mu(x_*(\epsilon))}{\partial
\epsilon}\approx
\frac{g(3\epsilon+2)}{12}\left(\frac{12}{3\epsilon+2g^2}\right)^{3/2}
-\frac{3g\epsilon(3\epsilon+4)}{16(3\epsilon+2g^2)}
\left(\frac{12}{3\epsilon+2g^2}\right)^{3/2}=0\,,
\end{equation}
and then
\begin{equation}
\epsilon_m=\frac{4}{3}g^2\frac{2}{1-2g^2+\sqrt{(1-2g^2)^2-4g^2}}\,.
\end{equation}

For a small $g$,
\begin{equation}
\Phi(0)_{*max}\approx\sqrt{\frac{4}{3}}g=1.15{\sqrt{1-q^2}}\,,
\end{equation}
\begin{equation}
M_{*max}=\frac{\mu(x_*(\epsilon_m))}{g}\approx\frac{4\sqrt{2}}{9g}=0.629\frac{1}{\sqrt{1-q^2}}\,,
\end{equation}
and
\begin{equation}
r_{*max}\equiv
\frac{x_*(\epsilon_m)}{g}\approx\frac{\sqrt{2}}{g}=1.41\frac{1}{\sqrt{1-q^2}}\,.
\end{equation}

We find that the approximation is qualitatively good, and the deviations are
slightly worse%
\footnote{In this scheme, the approximate values may always be larger than the
numerical values. Therefore, no purpose is served by making a further
correction due to the electric field.}
 than those in the previous approximation discussed in
Sec.~\ref{sec4}.

%%%%%%%%%%%%%%%%%%%%%%%%%%%%%%%%%%%%%%%%%%%%%%%%%%%%%%%%%%%%%%%%%%%%%%%%%%%
\acknowledgments
%%%%%%%%%%%%%%%%%%%%%%%%%%%%%%%%%%%%%%%%%%%%%%%%%%%%%%%%%%%%%%%%%%%%%%%%%%%
%Acknowledgements
%%%%%%%%%%%%%%%%%%%%%%%%%%%%%%%%%%%%%%%%%%%%%%%%%%%%%%%%%%%%%%%%%%%%%%%%%%%
%\begin{acknowledgments}
%We thank
%the organizers of JGRG21, where our
%partial result %({\tt [arXiv:10mm.xxxx]}) 
%was presented. %for elucidating comments.
%This study is supported in part by the Grant-in-Aid of Nikaido Research 
%Fund.
%\end{acknowledgments}
%%%%%%%%%%%%%%%%%%%%%%%%%%%%%%%%%%%%%%%%%%%%%%%%%%%%%%%%%%%%%%%%%%%%%%%%%%%

We thank Prof.~Kenji Sakamoto for much inspiration from his master's thesis
submitted about two decades ago.

%We would like to thank Editage (www.editage.jp) for English language editing.

%%%%%%%%%%%%%%%%%%%%%%%%%%%%%%%%%%%%%%%%%
%%%%%%%%%%%%%%%%%%%%%%%%%%%%%%%%%%%%%%%%%
%%%
%%%   References
%%%
%%%%%%%%%%%%%%%%%%%%%%%%%%%%%%%%%%%%%%%%%
%%%%%%%%%%%%%%%%%%%%%%%%%%%%%%%%%%%%%%%%%
%%%%%%%%%%%%%%%%%%%%%%%%%%%%%%%%%%%%%%%%%%%%%%%%%%%%%%%%%%%%%%%%%%%%%%%%%%%
%thebibliography
%%%%%%%%%%%%%%%%%%%%%%%%%%%%%%%%%%%%%%%%%%%%%%%%%%%%%%%%%%%%%%%%%%%%%%%%%%%
%\bibliographystyle{apsrev}
\bibliographystyle{apsrev4-1}
%\bibliography{}

%%%%%%%%%%%%%%%%%%%%%%%%%%%%%%%%%%%%%%%%%%%%%%%%%%%%%%%%%%%%%%%%%%%%%%%%%%%
%%%%%%%%%%%%%%%%%%%%%%%%%%%%%%%%%%%%%%%%%%%%%%%%%%%%%%%%%%%%%%%%%%%%%%%%%%%
%%%%%%%%%%%%%%%%%%%%%%%%%%%%%%%%%%%%%%%%%%%%%%%%%%%%%%%%%%%%%%%%%%%%%%%%%%%
\end{document}